\documentstyle[12pt]{article}
\input epsf.tex
\def\DESepsf(#1 width #2){\epsfxsize=#2 \epsfbox{#1}}
\begin{document}

\baselineskip=18pt

\begin{center}
{{\Large\bf Review from theoretician}\footnote[1]{Concluding 
remark presented
at the 7th RCNP International Workshop on Polarized $^3$He Beams
and Gas Targets and Their Application, Kobe, Jan. 20-24, 1997}}\\
\vspace{10mm}
T.\ Morii\footnote[2]{E-mail:morii@kobe-u.ac.jp,~~~Phone:+81-78-803-0917,
~~~Fax:+81-78-803-0261}\\
\vspace{5mm}
{\it Faculty of Human Development, Kobe University, Nada, Kobe 657,
Japan}
\end{center}

\begin{center}
ABSTRACT

\vspace{5mm}
\begin{minipage}{130 mm}
The talks presented at the conference are summarized from an angle 
of a particle theorist.  After presenting a personal 
impression on the conference,
particular emphasis is placed on the spin structure of 
nucleons and symmetry breaking test.

%\\
%PACS numbers: 13.85.Ni, 13.88.+e, 14.20.Dh, 14.40.Lb, 14.70.Dj 
\end{minipage}
\end{center}

\vspace{5mm}
\begin{flushleft}
{\bf Introduction}\\
\end{flushleft}
It is my great honor to be invited to give a review on this conference.
But before starting my talk, a few words seem to be necessary.  When I 
was asked by Prof. Tanaka, the chairman of this conference, 
to give a review on the conference, I could
not accept his request at first by saying, ``I am not an experimentalist and
have no good qualification to summarize the conference which will cover
an extremely wide range of subjects from quark to life, being mainly
experimental.''  But he persuaded me strongly by
saying, ``Don't worry.  I am just interested in your view on the 
conference and it is OK if you would cover only theoretical subjects
you are interested in.''

In any case, let me begin by apologizing to many of speakers for
neglecting their important reports.  Because of the limited time and
also according to the chairman's suggestion, after presenting 
my personal impression on the conference,
I am going to 
concentrate upon the subjects which I am interested in; 
spin structure of nucleons and symmetry breaking test.

%%%%%%%%%%%%%%%%%%%%%%%%%%%%%%%%%%%%%%%%%%%%%%%%%%%%%%%%%%%%%%%%%%%
\begin{figure}[htb]
\vspace{1 cm}
\centerline{ \DESepsf(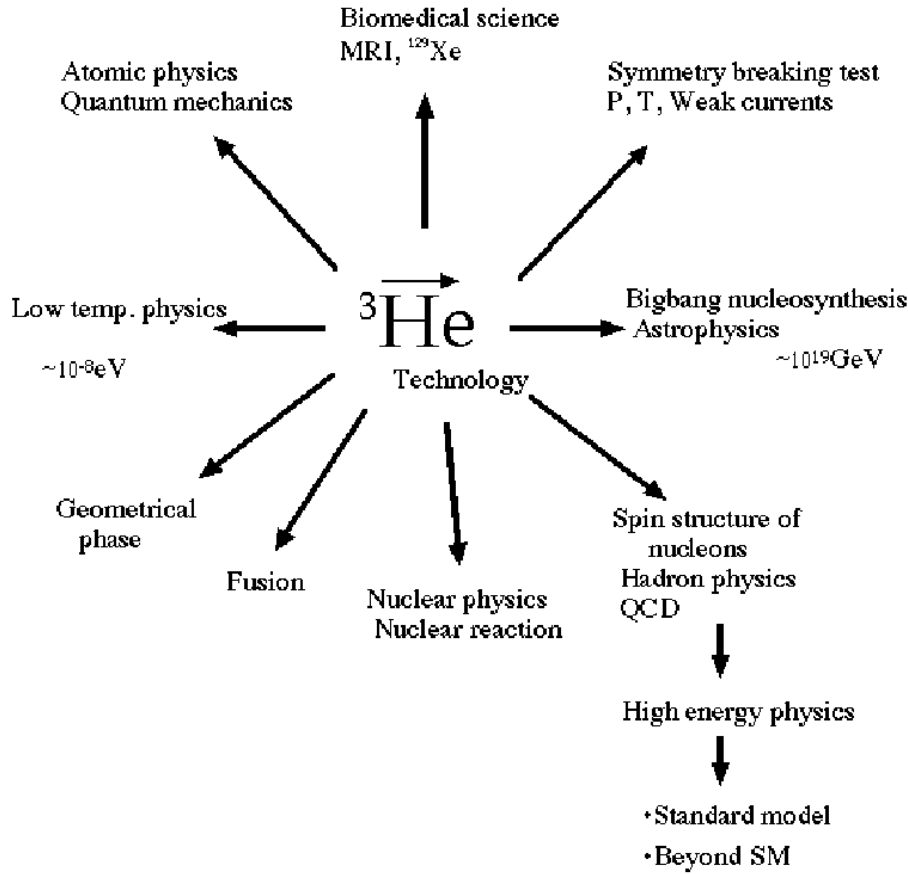 width 12 cm) }
\smallskip
\caption {from quark to life}
\end{figure}
%%%%%%%%%%%%%%%%%%%%%%%%%%%%%%%%%%%%%%%%%%%%%%%%%%%%%%%%%%%%%%%%%%% 

\begin{flushleft}
{\bf Personal impression on the conference}\\
\end{flushleft}
Before attending this conference, I thought that the polarized $^3$He must
be a very specified theme in nuclear physics.  But with all of the talks
presented here, I became aware that my first thought 
was completely wrong.  It is surprizing to know that in these 13 years after 
the previous conference at Princeton university in 1984, physics
and technology of the polarized $^3$He were greatly developed and 
so many subjects presented in 
this conference became related to the polarized $^3$He. 
The topics covered here 
are really wide as arranged in Fig. 1.  They range literally
``from quark to life'' as presented as a sub-title of this conference.  
Many reports were new and interesting for me; 
technology in polarized $^3$He,
spin structure of nucleons, symmetry breaking test, atomic physics,
astrophysics, biomedical science, fusion problems, etc.
Among many talks, here I would like to just mention only one report, the 
experiment for testing geometric quantum-phase by using NMR gyroscope
presented by W\"ackerle\cite{wac}, which I am personally interested in 
as a particle theorist since it is the subject 
related to the origin of the gauge
symmetry which is one of the most important principle in 
quantum field theories.
In any case, I learned a lot from many exciting 
reports, though, frankly speaking, it was rather hard for me to 
have to listen to all of the talks because I could not follow well the
technical details of many experimental reports.   

Active talks on experimental works reminded me some examples of
epoch--making experiments in the physics history which were touched
in some talks: One is the experiment
by Michelson and Morley in 1887 a long time ago. They tried to observe
an extremely small difference of time in which light runs along the ether
wind and perpendicularly to it.  The time difference is expected to be 
really tiny, 
$O(10^{-16}sec)$, in their experiment.  Even now, it must be incredible 
to try to observe it directly.  But inventing a fantastic technique called 
Michelson interferometry, they could observe it which was consistent
with zero.  The result was serious, becoming the ground of special 
relativity discovered by Einstein.  Another example is the Lamb 
shift experiment done by Lamb and Retherford in 1947.
Using a microwave technique developed during the World War II, they
found a contribution to a very small level splitting between the 
$2S$ and $2P_{1/2}$ states in a hydrogen.
Before this experiment, many theorists who were struggling with the 
problem of divergence in QED, 
have considered that QED might not be applied for high energies.
In those days, a rather strong pessimism about the description of QED
was widely spread among many theorists.
But the Lamb shift was just the one which could be predicted by QED.
Before long, QED came back to be reliable together with a discovery
of the renormalization method.  The Lamb shift  experiment gave rise to 
a phase transition from pessimism to optimism.

These examples tell us that good experiments bring about 
a revolution in physics.
I am sure that many of the experimental reports presented here are
very promising for a revolutional development of physics and its
application in the coming century.  Active flavor of many experimental 
reports will be seen in the proceedings.

\begin{flushleft}
{\bf Spin structure of nucleons}\\
\end{flushleft}
Triggered with the measurement of the polarized structure function of
protons, $g_1^p(x)$, by EMC in 1987, the spin structure of nucleons became one
of the most challenging topics in the particle and nuclear physics 
community\cite{Cheng}.
What is so interesting with the spin structure of nucleons?  So far, static
properties of nucleons have been explained successfully with the naive quark
model while high energy scatterings have been understood well with the parton 
model motivated by QCD.  However, the polarized experiments have produced
a breakdown of such a simple interpretation.  Many questions have been
brought out and are still
to be answered: where does the proton spin come from?,  why are $s$-quarks 
polarized negatively?,  what about gluons?, how does QCD work?, etc.
Now we have come to know that 
nucleons have much more fruitful structure than expected so far.
One of the important physics exists in the spin sum rule of nucleons,
\begin{equation}
\frac{1}{2}=\frac{1}{2}\Delta\Sigma+\Delta g+\langle L_Z\rangle_q+
\langle L_Z\rangle_g~,
\label{eqn:sumrule}
\end{equation}
which has never been seriously considered before 
observation of polarized structure functions of protons.
In this sum rule, questions are: How large is each component? and
What is the underlying dynamics of this sum rule?
Several talks were related to these questions.  A new test
of the gluon polarization was proposed by Yamanishi\cite{yam} 
in the analysis of 
testing a color-octet mechanism for small-$P_T$ $\psi '$ 
productions in polarized $pp$ scatterings, whose experiments 
could be carried out at RHIC in the future.  Titov discussed an 
interesting test of 
$s$-quark contents in a proton in the 
analysis for photoproductions of $\phi$ mesons\cite{titov}.  He has pointed 
out that the energy near 2GeV, which might be preferable for 
SPring-8 machine, must be adequate for detection of its effect.
In passing, recently Ji has proposed a possible way to measure
the orbital angular momentum of quarks in the nucleon 
in the process he calls as ``deeply--virtual'' Compton scatterings\cite{Ji}.  
In any case, to understand the underlying dynamics of
eq.(1) would open the door to a new field of hadron physics.

There were also several discussions on hadron structures and 
related topics. Toki discussed an importance of 
spin effects in hadronization processes\cite{Toki}.
Mizuno compared the spin structure of nucleons composed of quarks, to the
one of 3- and 2-body nucleon systems, i.e. $^3$He, $^3$H and 
D\cite{Mizuno}.
Precision measurement of the neutron electromagnetic form factor
with MIT-Bates and TJNAF experiments using polarized $^3$He which was 
discussed by Gao\cite{Gao} and Meziani\cite{Mezi}, respectively, 
is also important for 
understanding the nucleon structure.  Because of the lack of a free
neutron target, the neutron structure has not so precisely been observed
as the proton structure.  Polarized $^3$He is very effective for studying the 
form factor and spin structure of neutrons.  The precise data 
on the neutron form factor 
will largely contribute to our deep understanding of nucleon structures 
with QCD.

Remaining questions were also discussed in many talks; flavor separation
of parton distributions, test of GDH sum rule, 
neutron spin asymmetries at large $x$, higher twist effects,
spin test of the standard 
model and beyond, etc. Various experimental facilities such as TJNAF, MAMI,
SPring-8, HERA, RHIC, and so on, which will come out soon, will 
serve for studying these subjects.
Tests of the GDH sum rule which is due to the low energy theorem
and the unsubtracted dispersion relation for the crossing-odd amplitude,
$f_2$, of Compton scatterings, will be carried out at
TJNAF, MAMI and SPring-8.  
Furthermore, an interesting plan for studying the $Q^2$ dependence
of the generalized GDH sum rule is under way with these facilities.
These experiments will produce important data connecting nonperturbative
and perturbative regions of hadron dynamics and hence provide a 
crucial test of QCD. 
Meziani presented new experimental proposals at TJNAF with respect to 
the test of the GDH sum rule and related topics as well as the measurement
of neutron spin asymmetries\cite{Mezi}.
Higher twist effects which are important for giving
another test of QCD and deducing a 
reliable structure function, will be examined by HERMES group.
The present status and a rich program of HERMES experiments
were presented by De Schepper\cite{Schepp}.

\begin{flushleft}
{\bf Symmetry breaking test}\\
\end{flushleft}
Let me turn to the symmetry breaking test.  Why symmetry breaking tests
are so important?  It is because it can confirm or rule out the 
fundamental theory among many theories.  Whether people like a 
symmetric world or broken symmetric world belongs to their own personal 
favorite.  It is interesting to know that while many gardens in Europe
have beautiful geometrical symmetries, those in Japan have no such
symmetries and rather symmetries are broken there.  In any case, 
we are now living in the world
originated from the standard model with 
$G=SU(3)_c\times SU(2)_L\times U(1)_Y$.
The standard model is a local field theory based on three grounds; 
renormalizability, gauge principle and spontaneous symmetry breaking(SSB).  
Renormalizability is necessary to guarantee finite solutions.  
Gauge principle uniquely determines the interaction between matter and
gauge bosons.  As is well known, 
the fundamental theory has hardness in a sense that it has no room
for introducing arbitrariness by hand.  Renormalizability and 
gauge principle that all of the modern fundamental theories
should possess, are fundamental grounds which make the theory be hard
and beautiful.  On the other hand, 
the idea of SSB has not yet been tested 
in any field theories including the standard model, though it 
is also extremely important in many fields of
physics.    At present, one of
the most important job with the standard model is to test the 
mechanism of the SSB, i.e. the Higgs mechanism 
and find the Higgs boson.  Many theoretical predictions and 
phenomenological analyses in particle physics are 
also related to these topics.

In this conference, interesting discussions on symmetry breaking tests
were presented.  Detailed analyses on  weak currents which
can give a possible test of the standard model and beyond, 
were discussed by Govaerts\cite{Gov} and Souder\cite{Soud} 
with a new technique of
muon capture on $^3$He.  The most stringent limit of $T$--violation has
been given by the electric dipole moment of the neutron\cite{Gol}.
Masuda\cite{Masuda} discussed a rigorous 
test of $P$-- and $T$--invariance through
the neutron transmission experiment.  Special advantages of the neutron
transmission experiment were discussed in detail and 
the present limit of the electric dipole moment of 
the neutron was given.

\begin{flushleft}
{\bf Final words}
\end{flushleft}
In summary, the polarized $^3$He is fantastic matter which
is deeply related to so many topics spread from fundamental 
to applied sciences.
This conference is really very unique and not just a repeat of other 
conferences.  I hope that the technology and physics of the 
polarized $^3$He will be significantly developed toward the coming century
and the progress should be reported in the next HELION conference
planned in 2001.

%\begin{center}
%{\large \bf Figure captions}
%\end{center}
%\begin{description}
%\item[Fig.\ 1:] From quark to life
%\end{description}
\end{document}